\documentclass[conference]{IEEEtran}
\usepackage{graphicx}
\usepackage{url}

\begin{document}

\title{Displacement Is Endogenous: AI Workers in a Search-and-Matching Labor Market}

\author{\IEEEauthorblockN{Ibrahim Denis Fofanah}
\IEEEauthorblockA{Pace University\\
New York, NY, USA\\
\texttt{if57774n@pace.edu}}
}

\maketitle

\begin{abstract}
We build a labor market in which AI systems enter as workers: they apply for jobs, pass screenings, get hired, and get dismissed, alongside humans. Firms post vacancies, workers search with friction, and product demand slopes downward. Across 30 seeded runs per configuration, AI entry at 70\% of the human wage raises unemployment by 4.1 percentage points (95\% CI 3.5 to 4.6) and leaves 12.2 fewer humans employed (CI 10.5 to 13.9). Layoffs dominate when AI is much cheaper than humans: the layoff margin accounts for 7.2 of the lost jobs and reduced hiring for 5.0, though most dismissed workers find work again. A wider price gap reduces displacement, because cheap AI makes firms expand: at price parity, with zero dismissals, the unemployment effect is largest at 9.2 points (CI 8.6 to 9.9), all of it through hiring. That parity result survives cutting AI applications from five per period to one. Demand elasticity decides how many workers AI displaces. Screening changes who gets hired and leaves the displacement total unchanged.
\end{abstract}


\section{Introduction}

A 2026 survey of 500 US job seekers found that one in five of those using AI let it submit job applications automatically \cite{gdoc2026}. We model the limiting case of that trend: a labor market in which AI systems enter as workers. They apply for jobs, pass screenings, get hired, and get dismissed, alongside humans. In such a market, displacement is an equilibrium outcome of the same search-and-matching process that has always allocated workers to firms, with a new kind of applicant in the pool. The layoff is endogenous.

Whether one reads an AI hire as employment or as software procurement does not change the economics. A firm filling a position with an AI worker and a firm buying AI software are the same modeled event: a position filled at a per-period cost. The model takes no position on which description is correct.

We build this market in the Diamond-Mortensen-Pissarides tradition \cite{diamond1982,mortensen1994,nobel2010}. Firms face constant-elasticity product demand and choose how many positions to keep open. Humans find openings through search friction. AI workers face none and apply to several firms every period. Firms hire the cheapest adequate applicant per opening and dismiss a human when an adequate AI is available and the wage saving covers the firing cost. Wages are fixed.

Three results stand out. First, layoffs dominate when AI is much cheaper than humans: the layoff margin accounts for 7.2 of the 12.2 lost jobs and reduced hiring for 5.0. Most dismissed workers find work again, but the churn feeds the hiring competition that AI wins. As AI approaches price parity, dismissals stop and all displacement runs through hiring. Second, a wider price gap reduces displacement, because cheap AI makes firms expand and expansion is what keeps humans employed. Third, demand elasticity decides how many workers AI displaces, from 7.2 points of extra unemployment at low elasticity to near zero at high elasticity. A matching-function sweep moves the result only modestly: halving matching efficiency shifts displacement from 4.1 to 3.4 points and doubling it to 4.8, while the matching-elasticity sweep gives 4.0 to 4.9 points (Section V.G).

The parameters are stylized, so these results show a mechanism and put no numbers on real markets. What they show is that once AI systems compete as workers inside the matching process, displacement follows from the structure of that process at every price we tested.

\section{Related Work}

To our knowledge, no existing model lets AI systems compete as workers inside a search-and-matching labor market. The closest attempts put AI everywhere except the applicant pool. EconAgent \cite{econagent} simulates macroeconomic activity with LLM-empowered agents; the AgentTorch labor market model card \cite{agenttorch} documents a similar facility; the LLM Economist \cite{llmeconomist} and EconGym \cite{econgym} build richer economic environments; and recent work studies strategic behavior among competing AI agents directly \cite{chiu2026}. In all of them, AI reasons about the economy. In ours, AI applies for jobs in it.

The theoretical backbone is the DMP framework: Diamond's search equilibrium \cite{diamond1982}, the Mortensen-Pissarides model of job creation and destruction \cite{mortensen1994}, and the 2010 Nobel recognition of search-friction analysis \cite{nobel2010}. We keep the matching function and the entry margin but replace Nash bargaining with fixed wages, which isolates the displacement channel we study.

On the empirical side, the Iceberg Index \cite{iceberg} quantifies the technical exposure of skills to AI without measuring displacement, while Althoff and Reichardt \cite{althoff2026} model task-specific technical change in general equilibrium. Exposure becomes displacement only through the hiring process, which is the gap this paper addresses. Our screening module builds on the Artificial Frictional Unemployment framework \cite{fofanah2026a,fofanah2026b}, which quantifies how keyword-based resume screening falsely rejects qualified humans. Li \cite{li2026} comes closest to our method, modeling unemployment formation with LLM-based cognitive agents and deep reinforcement learning, but there AI drives the simulation rather than competing inside it as a worker. Generative agents research \cite{park2023} shows that LLM-driven agents can sustain coherent behavior in simulated roles; our AI workers are simpler, rule-based applicants behind an LLM-compatible interface, which keeps the mechanism transparent.

\section{Model}

\subsection{Agents and timing}

The market has 300 human job seekers and 25 firms, each starting with 8 positions. Each period: (1) AI workers arrive if enabled; (2) separations occur; (3) firms adjust position counts; (4) workers apply; (5) firms screen and hire; (6) firms replace humans with AI where profitable. Humans have skill in [0.3, 1.0] and reservation wages that drift down with unemployment duration. AI workers have capability in [0.55, 0.95] and a per-period operating cost.

\subsection{Search, matching, and the AI application advantage}

Humans find openings through a Cobb-Douglas matching function, $M = A_m U^{\alpha} V^{1-\alpha}$, with efficiency $A_m = 0.6$ and elasticity $\alpha = 0.5$. Each meeting produces one application. AI workers skip this entirely: every unplaced AI applies to several firms per period (five at the headline), reads every posting instantly, and never tires. The AI applicant pool is effectively inexhaustible, so rejecting one AI applicant changes nothing. Another AI applicant takes its place.

Firms hire the cheapest adequate applicant per opening. When a human and an AI applicant cost the same, the human wins: applications are processed in arrival order and the sort is stable, so ties break toward humans. The parity results below therefore understate AI's hiring advantage rather than flattering it.

\subsection{Product demand and vacancy creation}

Each firm faces constant-elasticity residual demand, $p(q) = D q^{-1/\epsilon}$ with $\epsilon > 1$. The marginal revenue of the $n$-th position is $MR(n) = R (n_0/n)^{1/\epsilon}$, normalized to $R = 1.2$ at the initial firm size $n_0 = 8$. Each period, a firm adds a position when the expected surplus of the next slot, fill probability times (marginal revenue minus expected hire cost), exceeds the vacancy cost of 0.05, and cuts a position after 6 periods of persistent unfilled openings. This vacancy rule approximates DMP entry without solving for equilibrium: it gives firms the entry margin the fixed-slot version of the model lacked, without claiming equilibrium uniqueness. Fixed wages keep the focus on quantities.

\subsection{The replacement margin}

A firm dismisses a human worker when an adequate AI worker is available and the wage saving over a 12-period horizon covers the firing cost: $(w - c_{AI})\,12 > F$, where $F$ is the firing cost. The headline firing cost is 2.0; we sweep 0.5, 2.0, and 8.0. Nobody programs aggregate displacement. It is the sum of these firm-level decisions.

\subsection{Implementation}

No LLM calls have been made. All experiments use a deterministic, rule-based backend behind an LLM-compatible decision interface, so a language model can later replace any decision rule without changing the market mechanics. Each configuration runs over 30 common random seeds, and every treatment effect reported is AI-minus-no-AI paired by seed with 95\% confidence intervals.

\section{Experiments}

We report six sweeps, all with free vacancy creation unless noted.

\textit{Demand elasticity.} $\epsilon$ in \{1.5, 2.0, 4.0, 8.0\}, headline 2.0. Higher $\epsilon$ means more elastic demand.

\textit{AI operating cost.} $c_{AI}$ in \{0.7, 0.8, 0.9, 1.0\} of the human wage, headline 0.7. This tests whether the price gap governs displacement.

\textit{Application volume.} AI applications per period in \{1, 2, 5, 10\}, at AI cost 0.7 and at price parity (1.0). This tests whether the parity result depends on the five-application choice.

\textit{Screening.} Keyword versus semantic screening, crossed with AI on/off; plus the AI false-negative sweep (0, 0.075, 0.15).

\textit{Matching.} Matching efficiency 0.3 and 1.2, and matching elasticity 0.3 and 0.7, at the headline configuration, to test whether the matching function drives the result.

\textit{Firing cost.} 0.5, 2.0, 8.0, at $\epsilon = 2.0$ and at $\epsilon = 1.5$, to decompose the layoff channel from the hiring channel. A fixed-slot contrast (vacancy creation off) provides a mechanical benchmark.

\section{Results}

\subsection{Headline: layoffs dominate when AI is cheap}

At the headline configuration ($\epsilon = 2.0$, keyword screening, AI at 70\% of the human wage), AI entry raises the unemployment rate by 4.1 points (95\% CI 3.5 to 4.6) and reduces human employment by 12.2 workers (CI 10.5 to 13.9), paired by seed against the no-AI baseline. Firms expand from 250 to 312 positions, and AI holds 34.5\% of employment at the end of the run (Table I, Fig. 1).

Firms dismiss about 22 workers per run. Of 21.5 dismissed workers, 17.6 are employed again by the end of the run, and 20.2, or 94.0\%, find work again at some point. The paired firing-cost contrasts decompose the 12.2 total. With dismissals banned (firing cost 8.0), AI entry still costs 5.0 humans their employment (CI 3.9 to 6.1): the hiring channel, operating through the openings AI wins. Allowing dismissals (firing cost 2.0) costs 7.2 more (CI 5.5 to 8.9): the layoff margin. Layoffs dominate when AI is much cheaper than humans. The margin exceeds the roughly 4 dismissed workers still out at the end, because the churn itself does damage: dismissed workers re-enter an applicant pool where AI applies several times as often and wins the openings.

The dynamics confirm it (Fig. 1). The unemployment gap opens at 8.3 points in the entry period (CI 7.5 to 9.0), peaks at 9.1 the period after, narrows to 2.9 by mid-run as dismissed workers rematch and firms expand, and ends 4.5 points above the no-AI path. The AI share of employment climbs steadily from 0 to 34.5\%.

\subsection{Three regimes of demand elasticity}

Demand elasticity decides how many workers AI displaces. Table I shows the sweep in three regimes: low ($\epsilon = 1.5$), intermediate ($\epsilon = 2.0$), and high ($\epsilon = 4.0$ and 8.0) elasticity.

In the low-elasticity regime, displacement is large: unemployment rises 7.2 points (CI 6.3 to 8.0) and 21.6 fewer humans work. The firing-cost decomposition shows both channels at work: the hiring channel costs 13.0 workers (CI 11.8 to 14.1) and the layoff margin 8.6 (CI 6.7 to 10.5). Firms dismiss 22 workers per run, but dismissals do not explain the loss one-for-one; most dismissed workers are rehired, and reduced hiring does the larger share even here.

In the intermediate regime, the headline, firms expand substantially and the layoff margin does the larger share, as shown above.

In the high-elasticity regime, displacement nearly vanishes: +0.7 points (CI 0.3 to 1.2) and +0.3 points (CI -0.1 to +0.7). Firms grow to 349--367 positions, AI hiring into new slots rises, and dismissals fall to 14 and then 9. When firms can grow, AI applicants are absorbed into new positions; when they cannot, unplaced AI piles up and firms substitute it for existing humans. Either way AI ends up doing about a third of the work. What changes is whether humans keep theirs.

\begin{table*}[t]
\caption{Demand-elasticity sweep, keyword screening, AI-minus-no-AI paired by seed (30 seeds).}
\label{tab:elasticity}
\centering
\small
\begin{tabular}{@{}lccccc@{}}
\hline
$\epsilon$ & $d_u$, pts (95\% CI) & $d_{emp}$ (95\% CI) & dismissals/run & base slots & AI share, end \\
\hline
1.5 & +7.2 (+6.3, +8.0) & -21.6 (-24.1, -19.0) & 22 & 245 & 0.36 \\
2.0 & +4.1 (+3.5, +4.6) & -12.2 (-13.9, -10.5) & 22 & 250 & 0.35 \\
4.0 & +0.7 (+0.3, +1.2) & -2.2 (-3.5, -1.0) & 14 & 263 & 0.32 \\
8.0 & +0.3 (-0.1, +0.7) & -0.8 (-2.0, +0.4) & 9 & 277 & 0.32 \\
\hline
\end{tabular}\\[4pt]
{\footnotesize Base slots are mean no-AI positions per run; AI-run slots are 291, 312, 349, and 367.}
\end{table*}

\subsection{A wider price gap reduces displacement}

The cost sweep reverses the naive story (Table II). Raising AI operating cost from 0.7 to 1.0 of the human wage increases the unemployment effect from 4.1 points (CI 3.5 to 4.6) to 9.2 points (CI 8.6 to 9.9), and the employment loss from 12.2 to 27.7 workers. Dismissals fall to zero at costs 0.9 and 1.0, since the wage saving no longer covers the firing cost. Yet displacement grows.

Cheap AI makes firms expand: 312 positions at cost 0.7 versus 277 at price parity, and expansion is what keeps humans employed, because new positions go to whoever applies. Expensive AI leaves firms small, and AI still captures the openings that exist, applying to five firms per period while humans manage roughly one meeting each. At exact price parity, with no cost advantage and zero dismissals, AI takes 38\% of end-of-run employment and displaces 27.7 humans purely through the application asymmetry. Humans win every direct cost tie; AI wins on volume.

\begin{table*}[t]
\caption{AI operating-cost sweep, $\epsilon = 2.0$, keyword screening, paired against no-AI baseline (30 seeds).}
\label{tab:cost}
\centering
\small
\begin{tabular}{@{}lccccc@{}}
\hline
AI cost & $d_u$, pts (95\% CI) & $d_{emp}$ (95\% CI) & layoffs/run & AI slots & AI share, end \\
\hline
0.7 $\times$ wage & +4.1 (+3.5, +4.6) & -12.2 (-13.9, -10.5) & 22 & 312 & 0.35 \\
0.8 $\times$ wage & +7.0 (+6.2, +7.7) & -20.9 (-23.2, -18.6) & 22 & 297 & 0.36 \\
0.9 $\times$ wage & +7.5 (+6.9, +8.0) & -22.4 (-24.1, -20.6) & 0 & 290 & 0.37 \\
1.0 $\times$ wage & +9.2 (+8.6, +9.9) & -27.7 (-29.7, -25.7) & 0 & 277 & 0.38 \\
\hline
\end{tabular}
\end{table*}

\subsection{Screening and the displacement total}

Keyword screening falsely rejects qualified humans, and the penalty is real: unemployment runs 0.9 points higher under keyword than semantic screening without AI (CI 0.5 to 1.2), and 1.0 point higher with AI (CI 0.5 to 1.6). But the difference-in-differences, the amplification of the screening penalty by AI entry, is +0.2 points with a confidence interval from -0.3 to +0.7 (Table III). Screening changes who gets hired. It leaves the displacement total unchanged. The AI false-negative sweep tells the same story: false-negative rates of 0, 0.075, and 0.15 for AI applicants move post-entry unemployment from 32.10 to 32.33 to 32.38 points, all within overlapping intervals. Rejecting an AI applicant just promotes the next one.

\begin{table}[t]
\caption{Screening effects, $\epsilon = 2.0$, paired by seed (30 seeds).}
\label{tab:screening}
\centering
\small
\begin{tabular}{@{}lc@{}}
\hline
contrast & $d_u$, pts (95\% CI) \\
\hline
keyword minus semantic, no AI & +0.9 (+0.5, +1.2) \\
keyword minus semantic, AI entry & +1.0 (+0.5, +1.6) \\
amplification (difference-in-differences) & +0.2 (-0.3, +0.7) \\
\hline
\end{tabular}
\end{table}

\subsection{The parity result does not depend on application volume}

Cutting AI applications per period from five to one leaves the parity result essentially unchanged: at AI cost 1.0, the unemployment effect is 8.5 points at one application per period (CI 7.8 to 9.1), 8.9 at two, 9.2 at five, and 9.2 at ten. AI's remaining advantages, no screening losses, no reservation wage, and an application every period without fatigue, decide the openings even when humans win every cost tie and apply just as often. At cost 0.7, application volume matters more: the effect falls from 5.8 points at one application per period (CI 5.1 to 6.4) to 4.3 at ten (CI 3.6 to 4.9), and dismissals fall from 40 to 20 per run, because faster hiring absorbs AI into firm expansion instead of substitution. The hiring channel scales with the application asymmetry, but the parity finding does not hang on the number five.

\subsection{Fixed positions: the mechanical benchmark}

With vacancy creation switched off, positions cannot adjust, and AI entry raises unemployment by 23.7 points (CI 23.2 to 24.2) under keyword screening. Every AI hire is then a human unhired by construction. The free-entry model exists to show how much of that mechanical displacement survives once firms can respond. The answer is about a sixth of it, and the surviving part is the endogenous part.

\subsection{Matching sensitivity}

Displacement survives across the matching-function range we tested (Table IV). Halving matching efficiency to 0.3 lowers the unemployment effect to 3.4 points; doubling it to 1.2 raises it to 4.8. Varying the matching elasticity from 0.3 to 0.7 moves the effect from 4.0 to 4.9 points. The matching function matters at the margin. It does not drive the result.

\begin{table*}[t]
\caption{Matching-function sensitivity, headline configuration, AI-minus-no-AI paired by seed (30 seeds).}
\label{tab:matching}
\centering
\small
\begin{tabular}{@{}lccc@{}}
\hline
params & $d_u$, pts (95\% CI) & $d_{emp}$ (95\% CI) & layoffs/run \\
\hline
baseline (eff 0.6, $\alpha$ 0.5) & +4.1 (+3.5, +4.6) & -12.2 (-13.9, -10.5) & 22 \\
efficiency 0.3 & +3.4 (+2.8, +3.9) & -10.1 (-11.7, -8.4) & 17 \\
efficiency 1.2 & +4.8 (+4.2, +5.4) & -14.4 (-16.2, -12.5) & 23 \\
elasticity 0.3 & +4.0 (+3.6, +4.5) & -12.0 (-13.4, -10.7) & 20 \\
elasticity 0.7 & +4.9 (+4.4, +5.4) & -14.7 (-16.1, -13.3) & 21 \\
\hline
\end{tabular}
\end{table*}

\begin{figure*}[t]
\centering
\includegraphics[width=\textwidth]{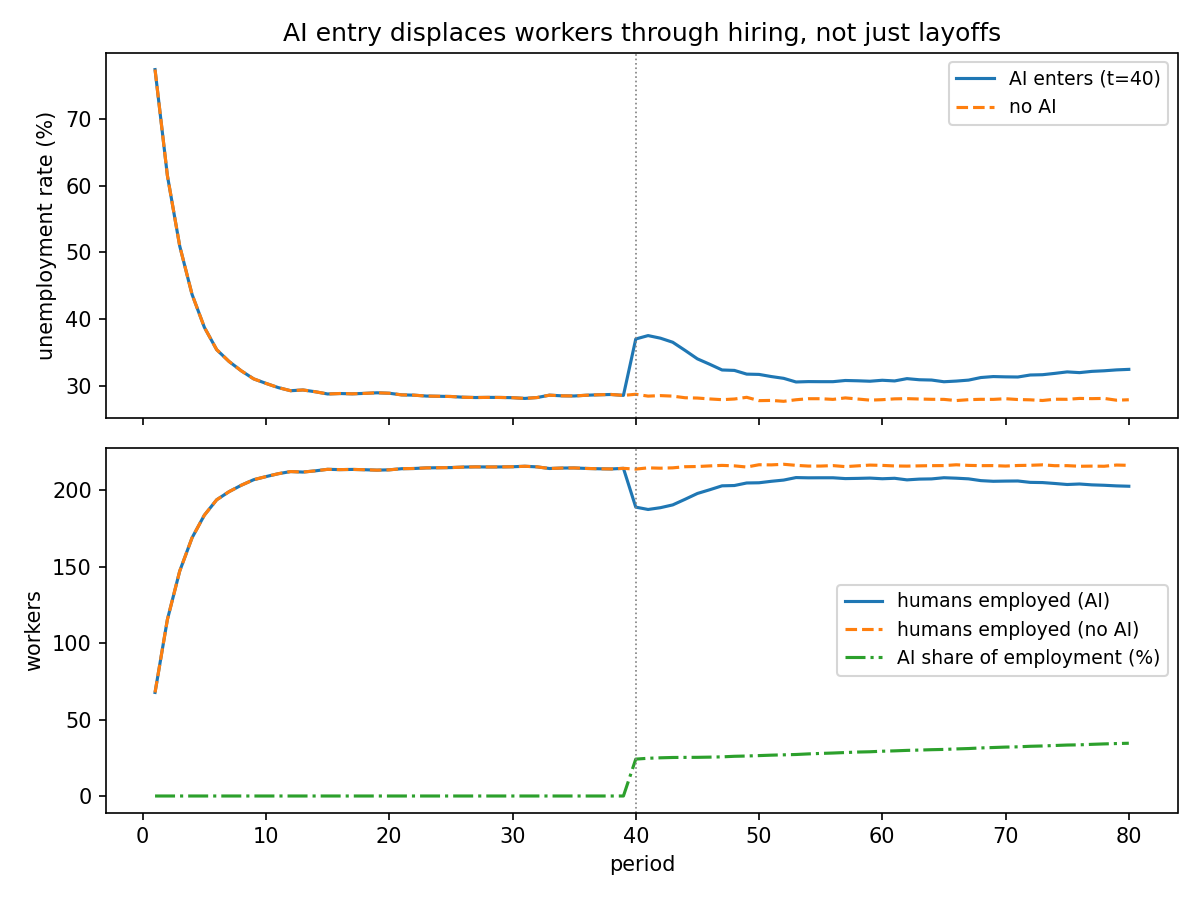}
\caption{Time series, headline configuration, averaged over 30 seeds. AI entry at period 40. Top: unemployment rate, AI versus no AI. Bottom: humans employed and AI share of employment. Human employment drops sharply at entry, recovers about half the loss, and settles persistently below the no-AI path while AI share climbs to 34.5\%.}
\label{fig:series}
\end{figure*}

\section{Limitations}

The parameters are stylized, so these results show a mechanism and put no numbers on real markets.

Wages are fixed by construction, so the model says nothing about wage adjustment cushioning displacement. The vacancy rule approximates DMP entry without solving for equilibrium. AI capability is a scalar, which flattens the task heterogeneity that exposure indices \cite{iceberg} measure. The five-applications-per-period choice is a stark representation of a real asymmetry, and the sweep in V.E shows the parity result is robust to it; at cost 0.7 the effect shrinks as applications rise. The tie-break favors humans, so the parity displacement is, if anything, understated. Finally, the backend is rule-based; replacing decision rules with language-model calls is future work, and it may change the quantitative results while leaving the mechanism intact.

\section{Conclusion}

AI entry displaces workers at every price we tested, including exact price parity. The displacement is endogenous to the matching process: AI applicants face no search friction, so they win the openings that exist, and product demand decides how many openings exist. Demand elasticity governs the magnitude, from 7.2 points of extra unemployment at low elasticity to near zero at high elasticity. Layoffs dominate when AI is much cheaper than humans; most dismissed workers find work again, but the churn feeds the hiring competition that AI wins. As AI approaches price parity, dismissals stop and the entire displacement runs through hiring.

Dismissal protections address the layoff margin, which is large when AI is cheap. But the hiring channel operates at every price, and at price parity it is the whole story. Either way AI ends up doing about a third of the work. What changes is whether humans keep theirs.

\end{document}